\documentclass[aps,prb,showpacs,twocolumn]{revtex4}
\usepackage{epsfig}
\usepackage{amsmath}

\begin{document}

\title{Friedel oscillations in disordered quantum wires: 
Influence of e-e interactions on the localization length}
\author{Y. Weiss, M. Goldstein and R. Berkovits}
\affiliation{The Minerva Center, Department of Physics, Bar-Ilan University,
  Ramat-Gan 52900, Israel}

\begin{abstract}

The Friedel oscillations caused due to an impurity located at one edge 
of a disordered interacting quantum wire are calculated numerically. 
The electron density in the system's ground state
is determined using the DMRG method,
and the Friedel oscillations data is extracted using the density difference
between the case in which the wire is coupled to an impurity and
the case where the impurity is uncoupled.
We show that the power law decay of the oscillations occurring for an interacting clean 
1D samples described by Luttinger liquid theory, is multiplied by an exponential
decay term due to the disorder.
Scaling of the average Friedel oscillations by this exponential term collapses the
disordered samples data on the clean results.
We show that the length scale governing the exponential decay may be
associated with the Anderson localization length and thus be used as a convenient
way to determine the dependence of the localization length on disorder and
interactions.
The localization length decreases as a function of the
interaction strength,
in accordance with previous predictions. 
\end{abstract}

%\pacs{73.21.Hb}{Quantum wires}
%\pacs{71.55.-i}{Impurity and defect levels}
%\pacs{71.55.Jv}{Disordered structures; amorphous and glassy solids}

\maketitle

\section{Introduction}

The interplay between repulsive interactions and disorder in low dimensional systems,
and their influence on the conductivity, were the subjects of many studies in recent years.
Some of this interest was
motivated by the experimental observations of a crossover in
the temperature dependence of the conductance of low density two dimensional
electrons from an insulating like dependence
at low densities to a metallic one at higher densities \cite{reviews}.
Nowadays it is generally accepted that even if such a 2D metal insulator
transition exists, it must be related to the spin degree of freedom \cite{punnoose05}
and therefore absent for spinless electrons.

It seems therefore clear that for spinless one-dimensional systems
no metal insulator transition is expected for repulsive interactions, although
for a certain range of attractive interactions a delocalized regime was found in
several studies \cite{schmitt1}.
Nevertheless, it was shown that there might be a certain strong disorder
and interaction regime, in which the localization length, or other
properties usually related to it such as the persistent current, 
increase \cite{abraham}. A sample dependent increase 
in the localization length was also reported for weaker values of disorder and
repulsive interactions for longer (of order of $100$ sites) wires\cite{pichard}.

On the other hand, several analytic studies \cite{giamarchi} have concluded 
that the localization length decreases monotonically with increasing 
repulsive interaction. Using either renormalization group \cite{apel_82} or
self consistent Hartree Fock \cite{suzumura_83} methods it was shown that 
the localization length, renormalized by the interaction, scales as
\begin{eqnarray} \label{eqn:xi_I}
\xi(g) \sim (\xi_0)^{1/(3-2g)},
\end{eqnarray} 
where $\xi_0$ is the localization length of the free electron system, and $g$ 
is the TLL (Tomonaga-Luttinger liquid)
interactions parameter with $g=1$ for non-interacting electrons.
Since for repulsive interactions $g$ decreases as a 
function of the interaction strength, one finds that the localization length 
always decreases as a function of the interaction strength.

One must be careful though to differentiate between weak and strong interaction
strength. A clean one dimensional system of spinless fermions on a lattice undergoes
a metal-insulator phase transition between a TLL 
and a charge density wave (CDW) as a function of the interaction strength.
This transition, caused by umklapp processes, is exhibited for commensurate fillings.
Once disorder is turned on, the TLL transport properties change drastically.
For more than a decade it is well known \cite{kane92} that the
conductivity of a TLL wire vanishes in the presence of impurities, 
thus a metal-insulator transition as a function of interaction strength no longer exists.
Yet, there is a difference between the two phases, since the TLL is replaced
by an Anderson insulator, while the CDW phase may remain a Mott type insulator, 
or become an Anderson insulator \cite{Pang93,giamarchi99}.

In this paper we investigate numerically the regime of the Anderson insulator caused by adding disorder
to the TLL phase. We study the effect of the interplay of weak interactions 
and disorder on the behavior of the Friedel oscillations in a wire due to its
coupling to an impurity at its edge.
Strictly speaking we probe the exponential decay of the Friedel oscillations
as a function of disorder and interactions, but for weak disorder this
decay length is equivalent to the localization length. 
It is important to note that the extraction of the localization length for
interacting systems is plagued with difficulties. The straightforward
method of measuring the decay length of the envelope of the single electron
state has no direct translation to a many electron state.
Nevertheless, one would prefer to stick to a ground state property of the
system, since the calculation of excited state dependent properties such as the 
conductance is computationally taxing. The sensitivity to boundary conditions
(i.e., persistent current) which is the natural candidate for a ground state property 
is problematic since it incorporates both interaction corrections to the localization
length as well as interaction corrections to the inverse compressibility of the
system \cite{berkovits96}.  
Thus, the study of the influence of interaction on the Friedel oscillation in the
Anderson phase is not only interesting on its own account, but it establishes a new
numerical method using a ground state property which is convenient for a 
direct evaluation of the localization length for not too strong disorder.
Using this method we show that the localization 
length as a function of the interaction strength decreases, 
in correspondence to Eq.~(\ref{eqn:xi_I}).

In general, the study of a dot (or impurity)
coupled to a one-dimensional lead, has been shown to shed some 
light over the physics of the lead. Certain thermodynamic observables, such as the 
occupation of the impurity level \cite{ours1,our_next,matveev}
and the corresponding electron density changes in the lead \cite{schmitt2,andergassen}, 
were recently used to analyze different wire properties, such as the strength and form 
of the interactions, and even the wire's phase (e.g., TLL vs. CDW). 
In a similar fashion, we show how the electron density of a disordered wire, 
coupled to an impurity level, can be used in order to extract its localization length.

Once a single-level impurity (dot) is coupled to a clean metallic system 
the density of electrons in its vicinity oscillates with a $2k_F$ period, and 
the envelope of the oscillations decays as a power law of $r$, the distance from the impurity \cite{friedel}.
For non-interacting systems the perturbation of the density in the vicinity of the
impurity depends on the dimensionality, $d$, of the system, and can be expressed as
\begin{eqnarray} \label{eqn:rho_frid}
\delta \rho(r) = A \frac {\cos(2k_F r+\eta)} {|r|^d},
\end{eqnarray}
where the coefficient $A$ and the phase shift $\eta$ do not depend on $r$. These oscillations
are the famous Friedel Oscillations (FO), which have been observed
experimentally during the last decade using various techniques, such as
scanning tunneling microscopy in low temperatures \cite{wielen}
and X-Ray diffraction \cite{brazovskii}.

Whereas for higher dimensions ($d \ge 2$) Eq.~(\ref{eqn:rho_frid}) is in 
general true even in the presence of interactions, this is not the case for 1D systems.
For the TLL phase, using field theoretical approaches, it was shown \cite{frid_LL}
that the $x^{-1}$ dependence is replaced by a different power law, $x^{-g}$.
For the non-interacting case $g=1$, it leads to the expected $x^{-1}$ decay,
while for repulsive interactions $g<1$ and thus a slower decay of the FO envelope
is expected.

From Eq.~(\ref{eqn:rho_frid}) it is clear that the observation of the density
fluctuations, either experimentally or numerically, is easier at short distances
in the vicinity of the impurity. 
When disorder is also introduced, this distance becomes
even shorter since there are also density fluctuations caused by the disorder.
Yet, in common experimental 1D situations disorder is usually present.
Therefore, although the presence of disorder hampers observing the FO, it
is beneficial to develop a method to tease the FO out of the density
fluctuations of a disordered system. 

The paper is organized as follows. In the following section 
we present the system's many particle Hamiltonian and the diagonalization method. We also
describe a simple method used to extract the FO data of disordered samples. 
In the current paper we restrict ourselves to the weak interactions regime (TLL), and
the results are presented in section $3$. Results for the CDW phase
(strong interactions), which show quite different physics, will be presented elsewhere \cite{cdw_next}.
In the last section we discuss the results mostly by a qualitative comparison to  
previous predictions, and offer some possible experimental realizations.

\section{Method}

\subsection{Hamiltonian and diagonalization method}

The system under investigation is composed of a spinless one dimensional electrons coupled to an impurity in 
one end. We model the one dimensional wire by a lattice
of size $L$ with repulsive nearest neighbor (NN) interactions and with an on-site disorder. 
The system's Hamiltonian is thus given by
\begin{eqnarray} \label{eqn:H_wire}
{\hat H_{wire}} &=& 
\displaystyle \sum_{j=1}^{L} \epsilon_j {\hat c}^{\dagger}_{j}{\hat c}_{j}
-t \displaystyle \sum_{j=1}^{L-1}({\hat c}^{\dagger}_{j}{\hat c}_{j+1} + h.c.) \\ \nonumber
&+& I \displaystyle \sum_{j=1}^{L-1}({\hat c}^{\dagger}_{j}{\hat c}_{j} - \frac{1}{2})
({\hat c}^{\dagger}_{j+1}{\hat c}_{j+1} - \frac{1}{2}),
\end{eqnarray}
where $\epsilon_j$ are the random on-site energies, taken from a uniform 
distribution in the range $[-W/2,W/2]$,
$I$ is the NN interaction strength ($I \ge 0$), and $t$ is the
hopping matrix element between NN, henceforth taken as unity.
${\hat c}_j^{\dagger}$ (${\hat c}_j$) is the creation (annihilation)
operator of a spinless electron at site $j$ in the wire, and
a positive background is included in the interaction term.

Without the disorder term, a similar system - in the limit $L \rightarrow \infty$ and with
periodic boundary conditions - has a well known exact solution. Depending 
on the interaction strength, the wire can be either metallic or insulating. The metallic
phase is described by TLL, occurring for $I<2t$, and the insulating
phase, in which $I>2t$, is a CDW. 
Previous studies which used wires of the order of a few hundreds sites have shown a
similar phase diagram \cite{ours1,our_next} even when employing open boundary conditions.
In order to stay in the TLL (Anderson insulator) regime for the clean (disordered) case,
we restrict ourselves to the range $0 \le I <2t$.

Introducing an impurity at one end of the wire results in adding 
the following term to the Hamiltonian:
\begin{eqnarray} \label{eqn:H_imp}
{\hat H_{imp}} &=& \epsilon_0 {\hat c}^{\dagger}_{0}{\hat c}_{0} 
-V ({\hat c}^{\dagger}_{0}{\hat c}_{1} + h.c.) \\ \nonumber
&+& I ({\hat c}^{\dagger}_{0}{\hat c}_{0} - \frac{1}{2})
({\hat c}^{\dagger}_{1}{\hat c}_{1} - \frac{1}{2}),
\end{eqnarray}
where $\epsilon_0$ describes the impurity strength, 
and $V$ is the hopping matrix element between the impurity and the first lead site.
Along this paper we use $\epsilon_0 \gg W$ and $V=t$.

The resulting Hamiltonian $\hat H = \hat H_{wire} + \hat H_{imp}$
describes a disordered one dimensional wire of length $L$ ($1 \le j \le L$), 
which is coupled to a single level at one of its edges ($j=0$).
Practically the $j=0$ site is equivalent to any other site, except for having
a constant onsite energy, whereas the other sites have energies drawn from a
distribution with a zero average over different realizations.

The Hamiltonian $\hat H$ was diagonalized using a finite-size DMRG method
\cite{white93,our_next}, and the occupation of the lattice sites were calculated,
for different values of $\epsilon_0$, $W$ and $I$. The size of the wire was up to $L=500$ sites.
During the renormalization process the number of particles in the system is not fixed, 
so that the results describe the experimentally realizable situation of a
finite section of a 1D wire which is coupled to a dot and to an external electron reservoir.

\subsection{Extracting the Friedel oscillations decay}

\begin{figure}[htb]\centering
\vskip 0.6truecm
\epsfxsize8.0cm
\epsfbox{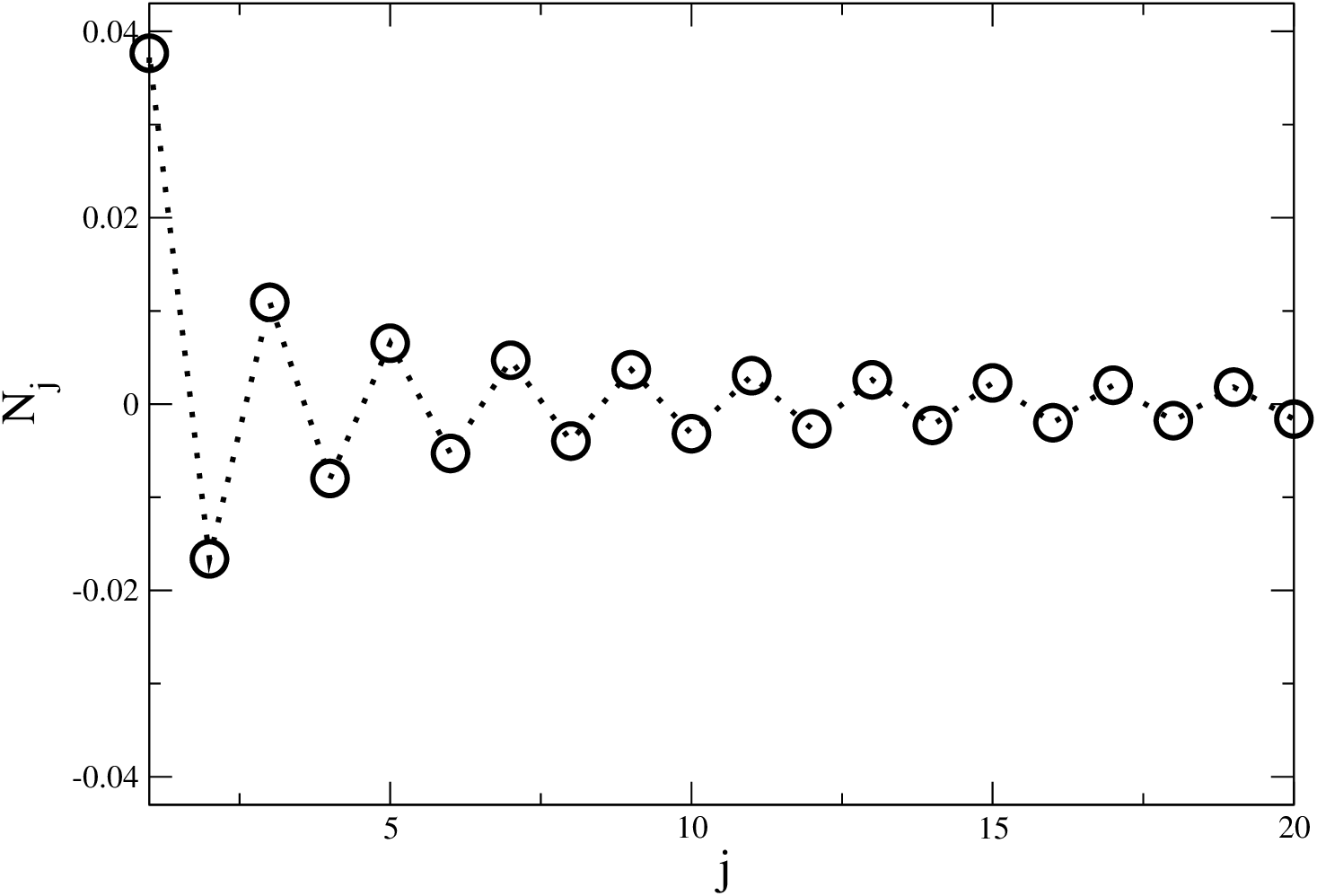}
\caption{\label{fig:fo_sample}
A typical form of the FO ($\epsilon_0=10$ and $L=280$) without disorder and
without interactions. The impurity is located at $j=0$ and the population 
of the first 20 lead sites is shown.
}
\end{figure}

When no disorder is present ($W=0$), $\hat H_{wire}$ has a
particle-hole symmetry, and the particle density of the wire's 
ground state is flat, with filling factor $n=1/2$. In this case $2k_F = \pi$ and
the oscillating part of Eq.~(\ref{eqn:rho_frid}) alternates according to $(-1)^j$.
Denoting by $n^{wire+imp}_j$ ($n^{wire}_j$) the electron density at site $j$ of the wire
when coupled (not coupled) to the dot, one has $n^{wire}_j=n=1/2$ for any $j$.
Clearly this is not the case in the presence of the impurity, and
the effect of the impurity is measured by $N_j \equiv n^{wire+imp}_j - n$.
A typical result of $N_j$, showing the $2k_F$ oscillations caused by the impurity at $j=0$,
is presented in Fig.~\ref{fig:fo_sample}. 

\begin{figure}[htb]\centering
\vskip 0.6truecm
\epsfxsize8.0cm
\epsfbox{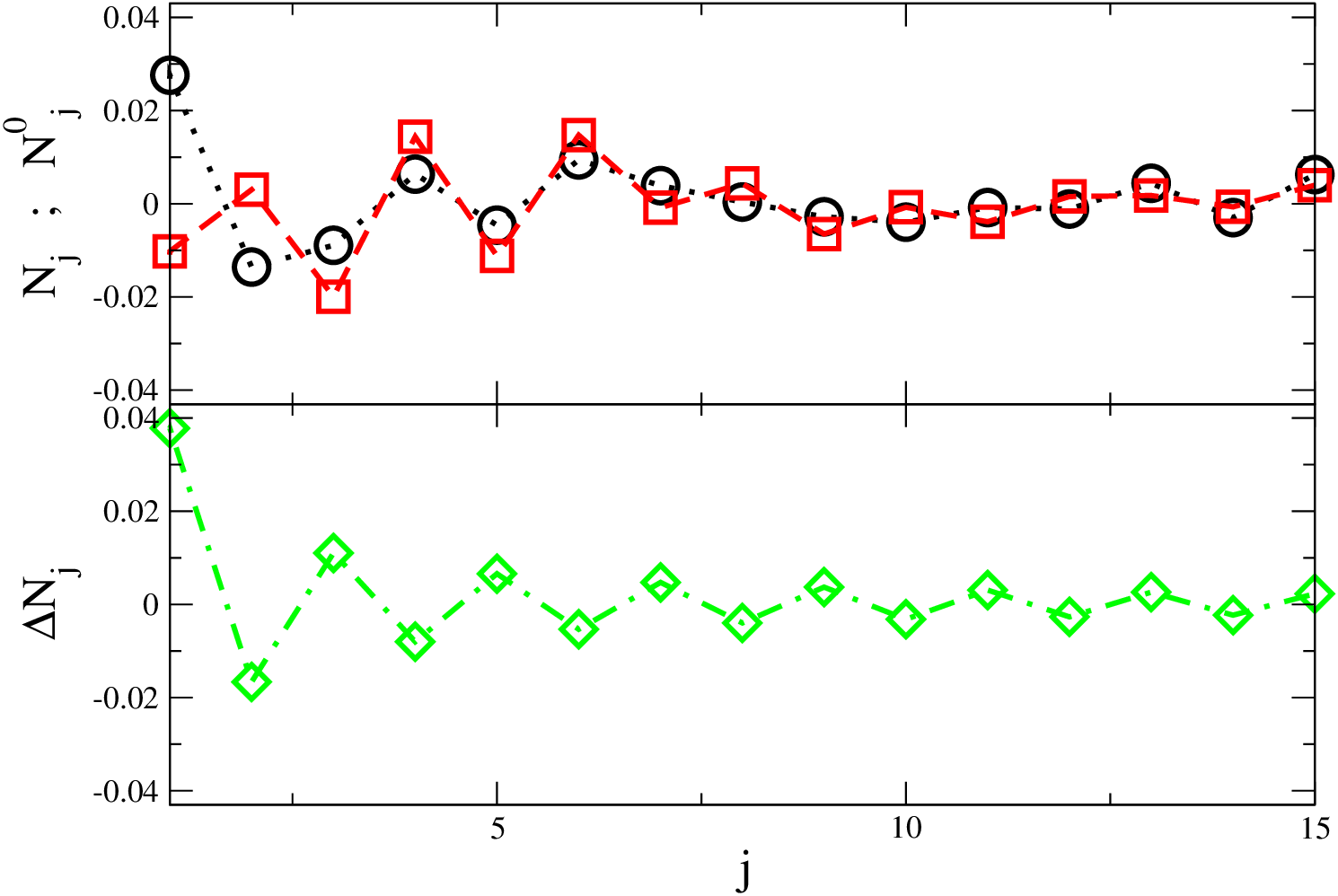}
\caption{\label{fig:frid_dis_sample}
A typical FO for a disordered sample with $L=280$, $W=0.1$ and $\epsilon_0=10$
(without interactions). 
The upper panel shows $N_j$ (circles) and $N^0_j$ (squares),
and the lower panel presents the difference between them ($\Delta N_j$).
The FO are observed much better using $\Delta N_j$ instead of $N_j$.
}
\end{figure}

When $W \ne 0$, on the other hand, although the average
filling factor is still $n \sim 1/2$,
there is no local symmetry between particles and holes, and 
the disorder effects are seen in the fluctuations of the electron density.
The density oscillations generated by the additional impurity are then difficult to discern,
since in a distance of a few lattice sites from the impurity the disorder fluctuations are dominant. 
A typical result of $N_j$ together with $N^0_j = n^{wire}_j - n$ 
(the electron density of the disordered wire without an impurity), 
is shown in the upper panel of Fig.~\ref{fig:frid_dis_sample}.

However, the influence of the impurity can be observed by isolating the density
fluctuations created by the disorder. This is achieved by 
comparing the electron density of the two cases shown in the upper panel of Fig.~\ref{fig:frid_dis_sample}, 
i.e. one with the additional impurity and the other without it, for every disorder realization.
Averaging over realizations is thus done for 
\begin{eqnarray} \label{eqn:Y_j_1}
\Delta N_j \equiv N_j - N^0_j = n^{wire+imp}_j - n^{wire}_j,
\end{eqnarray}
instead of just averaging over $N_j$. 
The curve of $\Delta N_j$ in the lower panel of Fig.~\ref{fig:frid_dis_sample} is for the 
same realization as in the upper panel.
It is obvious that the FO which were hardly seen for $N_j$ become clear 
once $\Delta N_j$ is considered.

\section{Results}

We begin with the results for a clean sample (i.e., $W=0$), so that the
density of the lead (disconnected from the impurity) is half everywhere. 
As was mentioned above, for a non-interacting wire ($I=0$ in our model), 
the decay of the oscillations is inversely proportional to the distance from the impurity,
and one expects to get, except for very short distances from the impurity,
$\Delta N_j = N_j - N^0_j = A (-1)^j j^{-1}$, 
where the amplitude $A$ does not depend on $j$.

\begin{figure}[htb]\centering
\vskip 0.6truecm
\epsfxsize8.0cm
\epsfbox{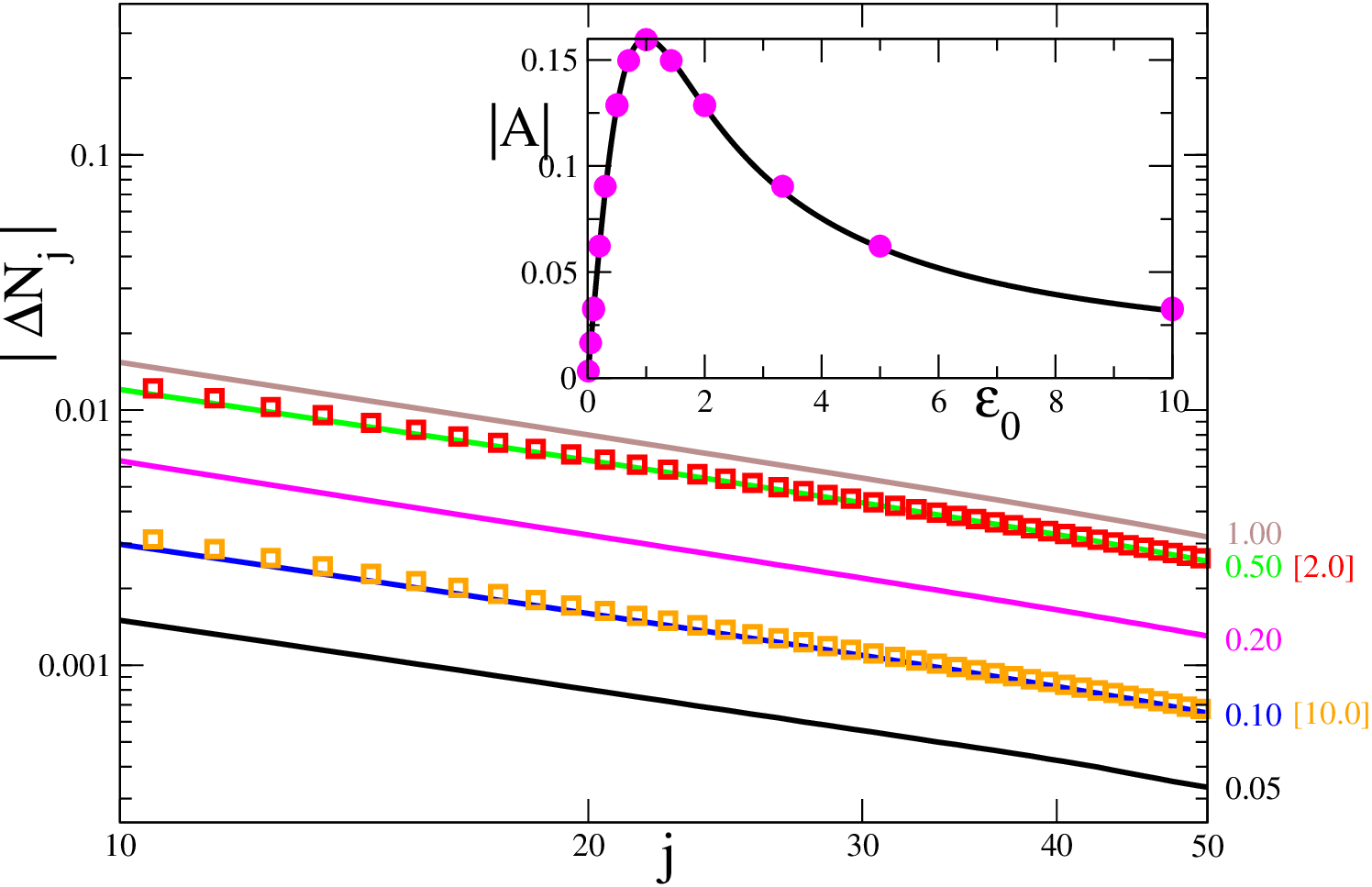}
\caption{\label{fig:A_E}
The FO decay (in log-log scale) for different values of $\epsilon_0$ (shown next
to the curves) for clean samples without interaction. 
The slope - representing the decay exponent  - is constant, 
and the only effect of $\epsilon_0$ is a change of the amplitude $A$.
The curves for $\epsilon_0=2.0,10.0$ are drawn with symbols.
Inset: the dependence of $A$ on $\epsilon_0$ for $V=t$
together with the exact formula Eq.~(\ref{eqn:A_exact}) 
which is derived in the appendix.
}
\end{figure}

However, the amplitude $A$ does depend on $\epsilon_0$, as presented in Fig.~\ref{fig:A_E}.
The limits of $\epsilon_0 \rightarrow 0$ and $\epsilon_0 \rightarrow \infty$ are well understood,
because in both of them the impurity does not play any role, the lead has a hard wall boundary,
and the particle-hole symmetry imposes that the FO amplitude goes to zero.
For finite values of $\epsilon_0$, the behavior of the amplitude is found to obey the relation
\begin{eqnarray} \label{eqn:A_exact}
A(\epsilon_0,V) = - \frac {1}{\pi} \big ( \frac {\epsilon_0 t}{V^2} + 
\frac {V^2}{\epsilon_0 t} \big ) ^{-1},
\end{eqnarray}
for which a complete proof is given in the appendix. The correspondence between the numerical 
results and this formula, as shown in the inset of Fig.~\ref{fig:A_E}, is excellent.

\begin{figure}[htb]\centering
\vskip 0.6truecm
\epsfxsize8.0cm
\epsfbox{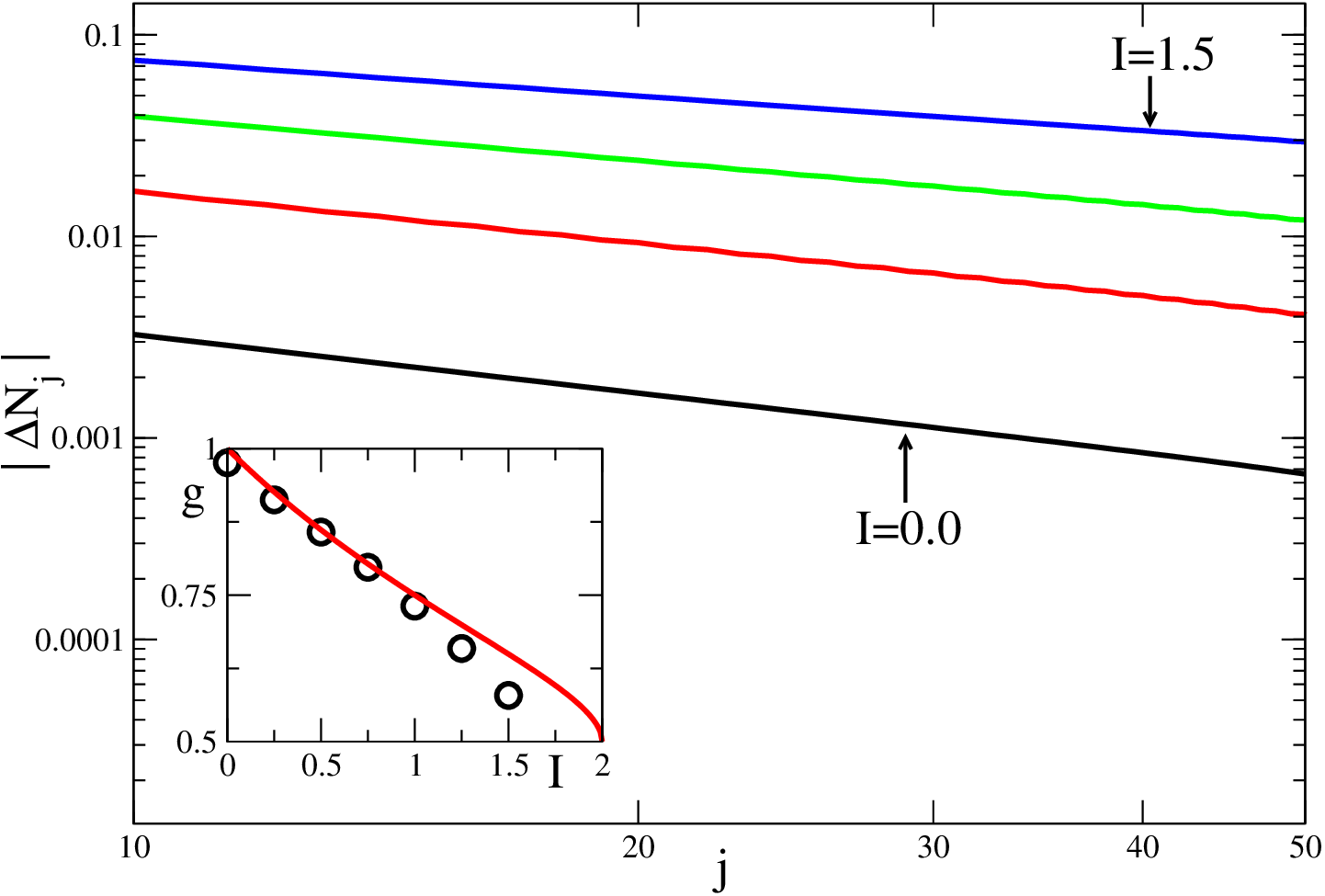}
\caption{\label{fig:I_slow_decay}
The DMRG results for the FO decay in log-log scale for $I=0,0.25,0.75$ and $1.5$ (bottom to top), 
with $\epsilon_0=10$ and $L=280$ and without disorder. As $I$ increases, the decay gets slower.
Inset: The interaction parameter g as found by fitting the FO decay to $x^{-g}$ (symbols),
together with the theory prediction Eq.~(\ref{eqn:g_theory}) (line). These results were
obtained by taking $L=500$ and $\epsilon_0=1$.
}
\end{figure}

We now move to the interacting case.
For $0 \le I<2t$, i.e. when the fermions in the lattice 
are described by the TLL theory, the decay is expected to be proportional \cite{frid_LL} 
to $j^{-g}$.
In our model the TLL parameter $g$ is given by \cite{g_formula}
\begin{eqnarray} \label{eqn:g_theory}
{g=\frac {\pi} {2 \cos ^{-1} [-I/(2t)]}}.
\end{eqnarray}
For non-interacting particles one gets $g=1$ so that the oscillations decay as $j^{-1}$, while in
the interacting regime a monotonic decrease of $g$ toward the limit $g=1/2$
occurs as a function of interaction strength. 
Thus, as $I$ becomes stronger, $g$ decreases, and a slower decay is predicted. This trend
is seen in the DMRG results presented in Fig.~\ref{fig:I_slow_decay}.

In the inset of Fig.~\ref{fig:I_slow_decay}, the results obtained for $g$
by fitting the FO decay of a $500$ sites wire, to the predicted decay of $x^{-g}$, 
are presented together with the theory prediction for $g(I)$ of Eq.~(\ref{eqn:g_theory}).
As can be seen, the results are in good accordance with the theory for
interaction strength $I/t \lesssim 1$. Similar results were obtained using 
other implementation of the DMRG method
(with a constant number of particles) \cite{schmitt2}, and by functional 
renormalization group studies \cite{andergassen}.
In these works it was argued that for the system sizes treated, the asymptotic regime 
in which the $x^{-g}$ behavior is predicted is not yet reached. In Ref.~\onlinecite{andergassen}
it was shown that using the fRG method, which is argued to be as accurate 
as the DMRG method, even $L$ of the order of $10^6$
is not sufficient to obtain the values of $g$ of Eq.~(\ref{eqn:g_theory})
for $I/t \gtrsim 1$.

\begin{figure}[htb]\centering
\vskip 0.6truecm
\epsfxsize8.0cm
\epsfbox{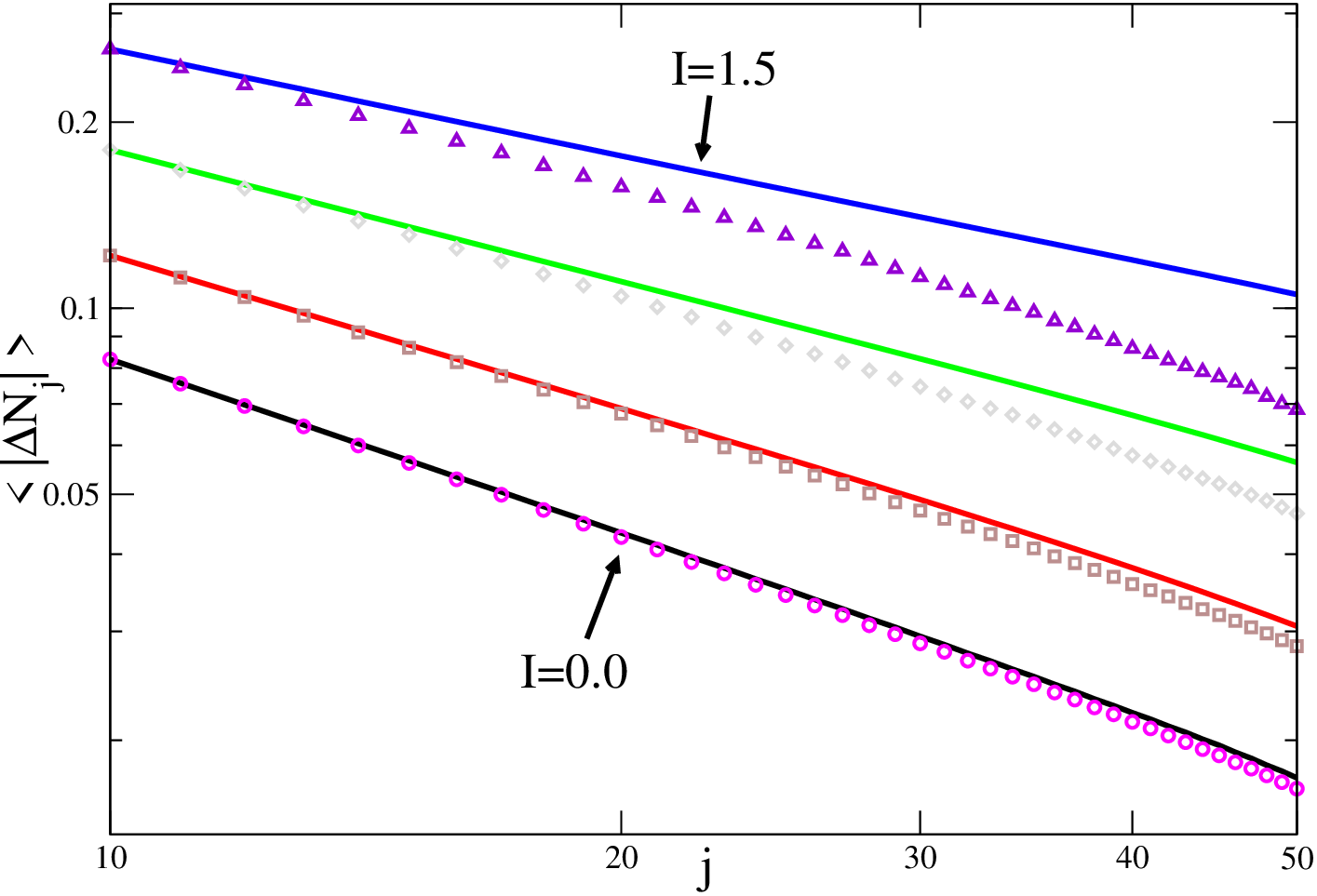}
\caption{\label{fig:decay_I}
The decay of FO for $I=0.0,0.5,1.0$ and $1.5$ (bottom to top)
with $L=500$ and $\epsilon_0=10$. 
The symbols are for $W=0.1$ and the lines are for the clean ($W=0$) case.
The disorder effect becomes significant for large values of $I$ where the
localization length is small. The average was done over 100 realizations.
}
\end{figure}

We now turn on the disorder by taking $W \ne 0$.
In this case the results of $\Delta N_j$ are averaged over 100 different realizations of disorder. 
In Fig.~\ref{fig:decay_I} the averaged particle density for $W=0.1$
is shown and compared to the $W=0$ case for various interaction strengths. 
As can be seen, for small values of the interaction the effect of disorder is very weak,
while for large values of $I$, the FO decays faster in the presence of disorder.
Zooming into these curves, it can be shown that the effect of disorder is 
to multiply the clean FO decay by an exponential factor $e^{-x/\xi}$, 
where $\xi$ is a characteristic decay length. 

\begin{figure}[htb]\centering
\vskip 0.6truecm
\epsfxsize8.0cm
\epsfbox{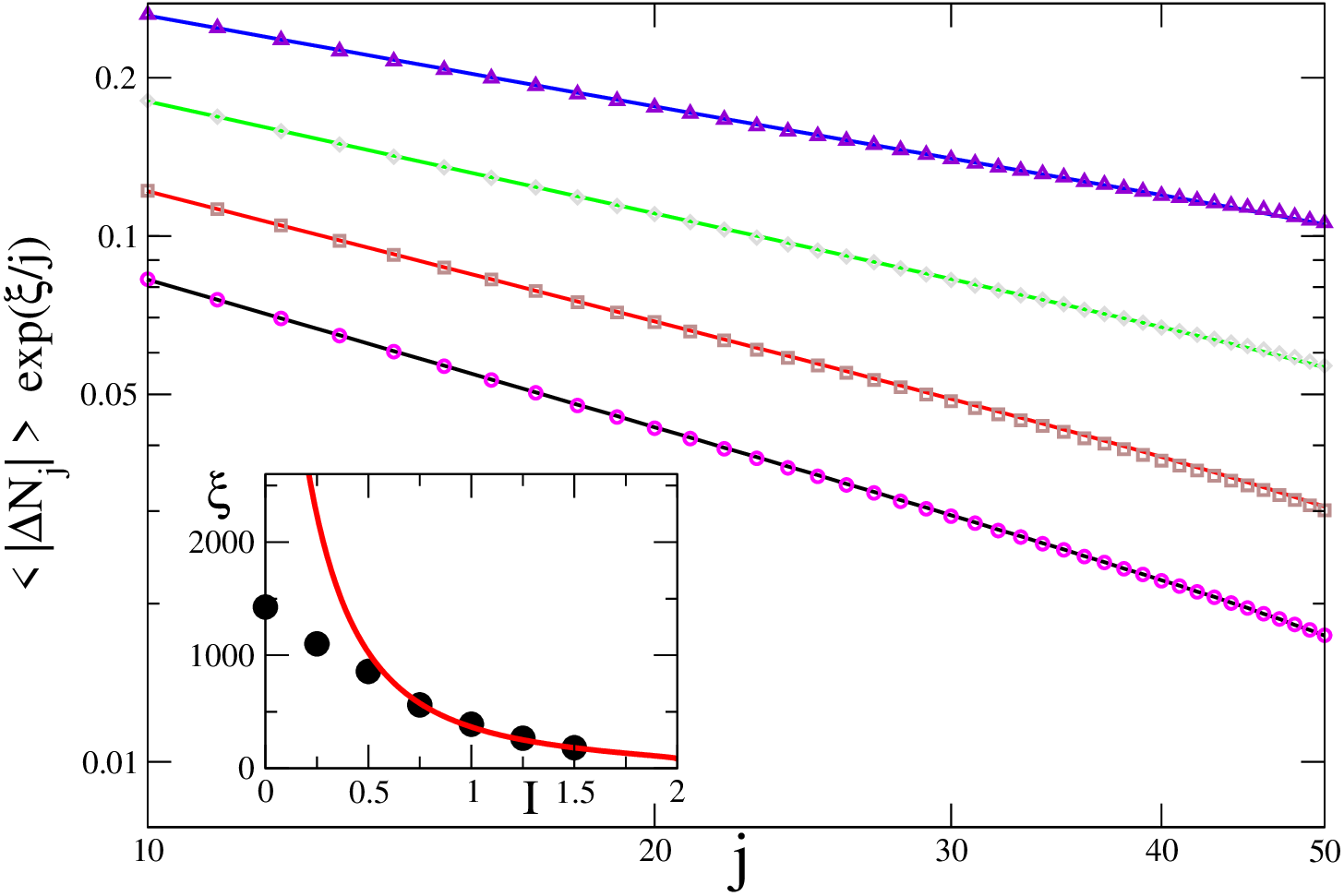}
\caption{\label{fig:LL_scaling_I}
The rescaled decay of FO for the $W=0.1$ curves over the clean curves with different interaction strengths. 
Inset: the localization length found by the best fit for each value of $I$ (symbols). 
The line corresponds to the theory prediction of Eq.~(\ref{eqn:xi_I}) with $\xi_0=7000$.
}
\end{figure}

For each strength of the interaction, 
one can rescale the disordered $W \ne 0$ curves, to the clean 
$W=0$ one by simply multiplying it by $e^{x / \xi}$, using $\xi$ as a fitting parameter. 
As can be seen in Fig.~\ref{fig:LL_scaling_I}, 
by using this rescaling method, the averaged disordered data collapses on the 
curves of the clean sample. 

The dependence of the decay length $\xi$ on the interaction strength $I$
is shown in the inset of Fig.~\ref{fig:LL_scaling_I}. 
We shall now show that this quantity
$\xi$ is effectively the mobility localization length.

The effect of disorder in the continuum limit can be divided to the 
forward and backward scattering terms.
Whereas the backward scattering term is related to the conductance and thus to
the localization length of the electrons, forward scattering processes
contribute only to the decay length of the FO, but not to localization. Thus, at first
sight $\xi$ does not necessarily correspond to the localization length. Nevertheless,
in this case one can argue that the contribution of the forward scattering
process to $\xi$ is small and therefore $\xi$ is a good measure of the localization
length.

Using standard bosonization technique it can be shown that the forward scattering 
processes result in the following term in the Hamiltonian:
\begin{eqnarray} \label{eqn:forward}
{H_{fs} = -\int {dx \eta(x)\frac {1} {\pi} \bigtriangledown \phi}},
\end{eqnarray}
where $\phi$ is the TLL field which is related to the density operator by 
$\rho(x) = -\frac{1}{\pi} \bigtriangledown \phi(x)$ and $\eta(x)$ is the $q \sim 0$ 
component of the random potential. Since the TLL Hamiltonian ($u$ being the velocity)
\begin{eqnarray} \label{eqn:H_LL}
{H_{TLL} = \frac {u} {2\pi} \int {dx [g (\bigtriangledown \phi)^2 +
\frac{1}{g} (\bigtriangledown \theta)^2}]},
\end{eqnarray}
depends on $\phi$ only through $(\bigtriangledown \phi(x))^2$, 
it is easy to show that by a redefinition of the field 
$\tilde{\phi}=\phi-\frac{g}{u}\int^x dy \eta(y)$ 
one can incorporate the $H_{fs}$ term inside $H_{TLL}$ and get a similar form of Hamiltonian.
Therefore, the forward scattering term is not expected to change the physics of the system.

Nevertheless, it was shown that this redefinition of the field has an effect on the
correlation functions \cite{giamarchi}. 
This results in a decay of the density-density correlation function,
which is, practically, the quantity we measure, and this decay
is not related to the conductance. It is an exponential decay of the form
$e^{-x/l}$, where $l = \frac {1}{2 D_f} (\frac{u}{g})^2$, 
and $D_f$ is the forward scattering strength of the disorder (defined in the non-interacting case).

For the decay described by the characteristic length $l$, one can find, using the
Bethe Ansatz solution, the factor $u/g$ for each value of $I$.
It is easy to show that $u/g$, and thus $l$, are monotonically increasing functions of $I$, 
as opposed to the FO decay length (see Fig.~\ref{fig:LL_scaling_I} in the inset).

Moreover, one can estimate $l$ quantitatively for the system we deal with.
The factor $u/g$ found from Bethe Ansatz solution 
ranges from $u/g=2$ for $I=0$ to $u/g \sim 4.5$ for $I=1.5t$. 
Denoting the amplitude of the disorder correlation function by $D$, 
i.e. $\langle V(x)V(x') \rangle = D\delta(x-x')$, one finds that
$D_f$ and $D_b$ (the forward and backward scattering disorder strengths) are 
of the same order of magnitude as $D$. 
For non-interacting spinless electrons in a one dimensional lattice \cite{shreiber}
$1/D_b \sim 100/W^2$. Substituting $W=0.1$, one gets $l$ of the order of $10^5$,
which is much longer than the observed decay length.

We thus conclude that the backward scattering processes are much more significant
in the model treated, thus $\xi$ is a very good approximation to the localization length, 
and its interaction dependence should be described by Eq.~(\ref{eqn:xi_I}).

Using the prediction of Eq.~(\ref{eqn:xi_I}) with the value
of the disorder we employ along this paper (order of $10^{-1}$), 
and recalling that without interactions $\xi_0 \sim 100/W^2$,
the localization length should range between $\xi(I=0) \sim 10^4$, which is much larger 
than the lattice sizes we considered, and thus almost doesn't influence the electron density, to $\xi(I=2) \sim 10^2$, 
in which the disorder effect should indeed be much more dominant, in agreement with the 
qualitative results presented in Fig.~\ref{fig:decay_I}.

The quantitative data (Fig.~\ref{fig:LL_scaling_I} (inset)) fits 
the theoretical predictions of Eq.~(\ref{eqn:xi_I})
for not too weak interactions. For weak interactions ($I \lesssim 0.5$)
no such fit was found, which is however expected, since for this regime
the theoretical localization length is much larger than the wire length.
The fact that the best fit to Eq.~(\ref{eqn:xi_I}) was for $\xi_0 \sim 7000$
(not the expected $\xi_0 \sim 10000$)
can be attributed to the same reason, as well as to the neglected forward scattering 
term which is stronger for weak $I$.
We also note that the exact choice of the wire slices over which the fit is done, can
change slightly the values of $\xi$. This, however, does not change the qualitative results,
showing monotonic decrease of $\xi$ as a function of the interaction strength.

To summarize, 
the effect of disorder on the FO decay in the Anderson regime can be described by an
extra exponential decay of the FO, which depends on the localization length,
of the form
\begin{eqnarray} \label{eqn:Y_j_2}
<\Delta N_j> = A (-1)^j j^{-g} \exp ( -j/\xi(g)),
\end{eqnarray}
where the localization length $\xi(g)$ decreases monotonically
as the interactions increase.

\section{Discussion and conclusions}

In conclusion, we have shown that the FO envelope,
in the TLL phase and the resulting disordered Anderson insulator phase,
is affected by both the interaction strength, and the disorder strength. 
Interactions actually enhances the effect of FO since it drops with a 
weaker power law $j^{-g}$, while disorder decreases the FO
oscillation since it adds an exponential factor to its decay form. 
We have shown that the length scale for this exponential decay 
is a good approximation to the mobility localization length, since it is weakly
influenced by forward scattering processes for weak disorder.
Thus we established a convenient way to evaluate the dependence of the
localization length on disorder and interaction using only
the ground state properties of the system. Qualitatively, the
localization length as a function of interaction for a given weak disorder
always decreases. As long as the localization
length is not much longer than the wire's length the localization length behavior
is quantitatively described by the renormalization group results\cite{apel_82}.
We have also analytically described the dependence of the FO amplitude on
the impurity strength.

Finally we remark to the experimental relevance of this work. The theoretical
treatment of disorder usually involves statistics over an ensemble of
many samples which is usually hard to obtain experimentally.
Furthermore, in the case we deal here, a measurement of FO on a disordered sample 
seems at first sight daunting. However,
the simple method we suggest in order to deal with the disorder, is 
in principle experimentally 
feasible, and solves these two difficulties.

Once a technical method for measuring the electron density is established,
it should be used twice for each sample, before and after the coupling of the
wire to the dot. In principle, by using a gate it should be possible to eliminate
the coupling between the dot and the wire.
Our results, as can be seen in Fig.~\ref{fig:frid_dis_sample}, 
which presents typical results for a particular realization,
point out that the difference between these two measurements 
should show a very clear FO, even for a specific sample.

\appendix*
\section{Friedel Oscillations in the 1D Tight-Binding model}

In this appendix we calculate 
$N(m)$, the density of electrons in site $m$, of a half filled one dimensional 
Tight-Binding lead, which is coupled to an impurity, in the asymptotic 
($m \gg 1$) limit. The system is described by the Hamiltonian
\begin{eqnarray} \label{eqn:H_app}
{\hat H} = 
\displaystyle 
\epsilon_0 {\hat c}^{\dagger}_{0}{\hat c}_{0} 
-V ({\hat c}^{\dagger}_{0}{\hat c}_{1} + h.c.) \\ \nonumber
-t \displaystyle \sum_{j=1}^{L-1}({\hat c}^{\dagger}_{j}{\hat c}_{j+1} + h.c.).
\end{eqnarray}

$N(m)$ can be calculated using the
retarded Green function of an electron in the m'th site $G^R (\omega;m,m)$
\cite{mahan}, and the relation (for a half filled band)
\begin{eqnarray} \label{eqn:N_G}
N(m) = - \frac {1}{\pi} \Im \int_{-2t}^0 G^R(\omega;m,m)d\omega,
\end{eqnarray}
where we are possibly neglecting bound states with energy lower than $-2t$, which give exponentially
small contributions for large m.
The Green function itself is determined by 
\begin{eqnarray} \label{eqn:G_mm}
G^R(\omega;m,m) &=& G_0^R(\omega;m,m) + \\ \nonumber
 G_0^R(\omega;m,1) ~~ \cdot &V& \cdot ~~ G(\omega;0,0) ~ \cdot ~ V ~ \cdot ~ G_0^R(\omega;1,m).
\end{eqnarray}

In this expression $G_0^R(\omega;m,l)$ is the bare (i.e., without dot) 
lead Green function, while
$G(\omega;0,0)=(\omega-\epsilon_0-\Sigma(\omega))^{-1}$ is the dot's Green function,
where $\Sigma(\omega) = \frac {V^2}{t} ( \frac {\omega}{2t} - i \sqrt{1-(\frac {\omega}{2t})^2})$
is the self energy of the dot \cite{ours1}. 
The first term in the RHS of the equation simply gives the constant $n=1/2$ occupation in the 
absence of the dot.

Substituting the known wave functions and energies of the tight-binding Hamiltonian
one finds
\begin{eqnarray} \label{eqn:G0_ml}
G_0^R(\omega;m,l) &=& \\ \nonumber
\frac {1}{L} &\displaystyle \sum _{k > 0}& \frac {\cos(ka(m-l)) - \cos(ka(m+l))} {\omega+2t \cos(ka)},
\end{eqnarray}
where $k=\frac{\pi}{L}n_k$, for integer $n_k$.
Transforming to integration over unit circle in the complex plane leads to
\begin{eqnarray} \label{eqn:G0_m1}
G_0^R(\omega;m,1) = -\frac{1}{t} \big{[} -\frac{\omega}{2t}+i \sqrt{1-(\frac{\omega}{2t})^2} \big{]} ^m.
\end{eqnarray}

Combining Eqs.~(\ref{eqn:N_G}), (\ref{eqn:G_mm}) and (\ref{eqn:G0_m1}), one can get,
\begin{eqnarray} \label{eqn:DN_m}
\Delta N(m) &=& N(m) - 1/2 = \\ \nonumber
-\frac{V^2}{\pi t^2} \Im \int_{-2t}^0 &d\omega&
\frac {\big{(} -\frac{\omega}{2t}+i \sqrt{1-(\frac{\omega}{2t})^2} \big{)} ^{2m} }
{ \omega - \epsilon_0 - \frac{V^2}{t} 
\big{(} \frac{\omega}{2t} - i \sqrt{1-(\frac{\omega}{2t})^2} \big{)} },
\end{eqnarray}
and by substituting $\omega = -2t \cos \theta$, we find
\begin{eqnarray} \label{eqn:DN_m2}
\Delta N(m) = \frac{V^2}{\pi t i} \int_{-\pi/2}^{\pi/2} d\theta
\frac {\sin(\theta) e^{i2m\theta}}
{2t \cos (\theta) + \epsilon_0 - \frac{V^2}{t} e^{i\theta}}.
\end{eqnarray}

\begin{figure}[htb]\centering
\vskip 0.6truecm
\epsfxsize3.0cm
\epsfbox{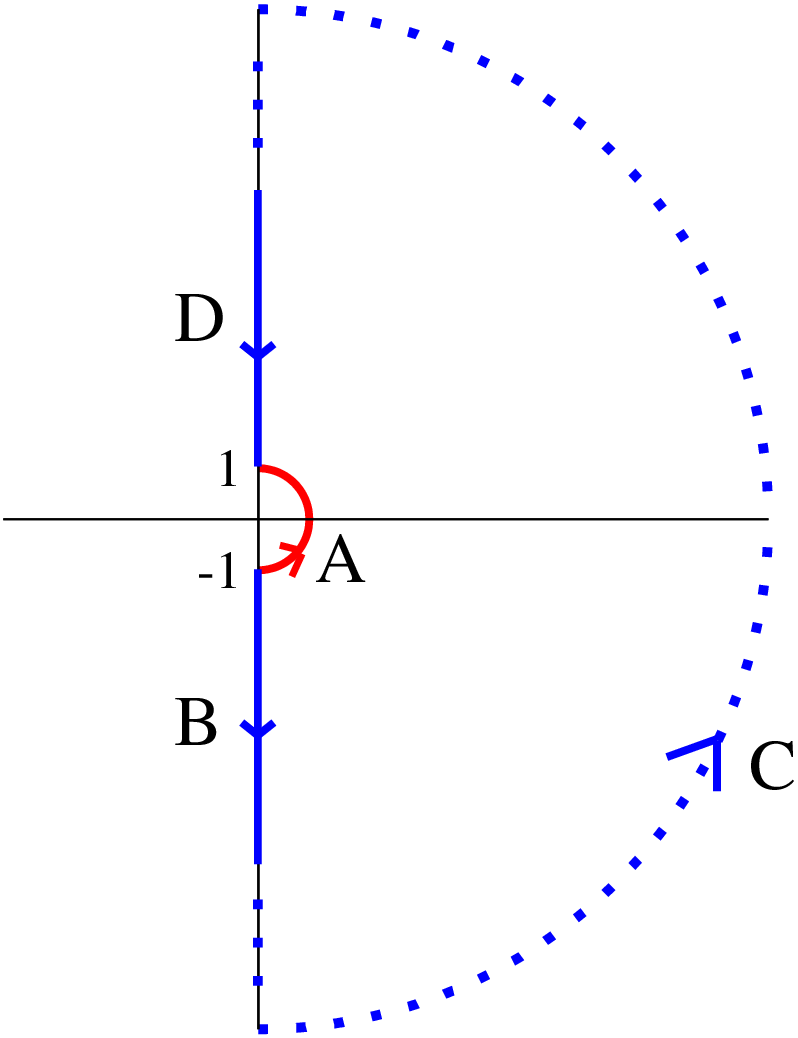}
\caption{\label{fig:contour}
The integration contours A and B-C-D which connect the points $[0,-1]$ and $[0,1]$.
}
\end{figure}

One now defines $z=e^{-i\theta}$ in order to get 
\begin{eqnarray} \label{eqn:DN_m3}
\Delta N(m) = -\frac{V^2 i}{2 \pi t^2} \int_A \frac {dz}{z}
\frac {(z^2-1)z^{-2m}}
{z^2+\epsilon_0 z/t + 1-V^2/t^2},
\end{eqnarray}
where the integration is over the right half of the unit circle, between the points 
$\pm 1$ on the imaginary axis (contour A in Fig.~\ref{fig:contour}).

Next we deform our contour to the contour B-C-D in Fig.~\ref{fig:contour}.
In doing so we neglect the contribution of poles which may occur inside the closed 
line A-B-C-D. These represent states bound at the impurity, and as we have mentioned above,
contribute exponentially small terms for large $m$.
The integration in parts B and D is done by defining $z=\pm ix$, respectively, $x \in [1,\infty)$,
while the contribution of the semicircle C vanishes as its radius goes to infinity.
Therefore we get
\begin{eqnarray} \label{eqn:DN_m4}
\Delta N(m) &=& \\ \nonumber
\frac{V^2}{\pi t^2} (-1)^m &\Im& \int_1^\infty 
\frac {(x^2+1)}
{x^2+i\epsilon_0 x/t - 1+V^2/t^2}
x^{-2m-1} dx.
\end{eqnarray}

For $m \gg 1$ the term $x^{-2m-1}$ varies much faster than the other terms,
and the rest of the integrand can be
evaluated at $x \sim 1$ to give $\frac {2}{V^2/t^2+i\epsilon_0/t}$. One thus gets the final form 
\begin{eqnarray} \label{eqn:N_j_exact2}
\Delta N(m) = \frac {(-1)^{m+1}}{\pi m} \big ( \frac {\epsilon_0 t}{V^2} + \frac {V^2}{\epsilon_0 t} \big ) ^{-1}.
\end{eqnarray}

\acknowledgments

We thank B. L. Altshuler, C. Lewenkopf and V. Fal'ko
for very useful discussions and the Israel Academy of Science 
(Grant 877/04) for financial support.


\begin{thebibliography}{2}

\bibitem{reviews} For reviews, see E. Abrahams, S. V. Kravchenko, and
M. P. Sarachik, Rev. Mod. Phys. {\bf 73}, 251 (2001); S. V.
Kravchenko and M. P. Sarachik, Rep. Prog. Phys. {\bf 67},
1 (2004); A. A. Shashkin, Physics-Uspekhi {\bf 48}, 129 (2005).

\bibitem{punnoose05} A. Punnoose and A. M. Finkel'stein, Science {\bf 310}, 289 (2005).

\bibitem{schmitt1} P. Schmitteckert, T. Schulze, C. Schuster, P. Schwab and U. Eckern, Phys. Rev. Lett. {\bf 80}, 560 (1998);
J. M. Carter and A. MacKinnon, Physical Review B {\bf 72}, 024208 (2005)

\bibitem{abraham} M. Abraham and R. Berkovits, Phys. Rev. Lett. {\bf 70}, 1509 (1993).

\bibitem{pichard} P. Schmitteckert, R. A. Jalabert, D. Weinmann and J. L. Pichard , Phys. Rev. Lett. {\bf 81}, 2308 (1998).

\bibitem{giamarchi}  T. Giamarchi, {\it Quantum Physics in One Dimension} (Oxford University Press, New York, 2003).

\bibitem{apel_82} W. Apel, J. Phys. C {\bf 15}, 1973 (1982).

\bibitem{suzumura_83} Y. Suzumura and H. Fukuyama, J. Phys. Soc. Jpn. {\bf 52}, 2870 (1983).

\bibitem{kane92} C. L. Kane and M. P. A. Fisher, Phys. Rev. Lett. {\bf 68}, 1220 (1992);
Phys. Rev. B {\bf 46}, 7268 (1992); Phys. Rev. B {\bf 46}, 15233 (1992).

\bibitem{Pang93} H. Pang, S. Liang, J. F. Annett, Phys. Rev. Lett. {\bf 71}, 4377 (1993).

\bibitem{giamarchi99} E. Orignac, T. Giamarchi, and P. Le Doussal, Phys. Rev. Lett. 
{\bf 83}, 2378 (1999).

\bibitem{berkovits96} R. Berkovits and Y. Avishai, Phys. Rev. Lett. {\bf 76}, 291 (1996).

\bibitem{ours1} M. Sade, Y. Weiss, M. Goldstein and R. Berkovits,
Phys. Rev. B {\bf  71}, 153301 (2005).

\bibitem{our_next} Y. Weiss, M. Goldstein and R. Berkovits,
cond-mat/0610543.

\bibitem{matveev} A. Furusaki and K. A. Matveev, Phys. Rev. Lett. {\bf 88}, 226404 (2002).

\bibitem{schmitt2}  P. Schmitteckert and U. Eckern, Phys. Rev. B. {\bf 53}, 15397 (1996).

\bibitem{andergassen}  S. Andergassen, T. Enss, V. Meden, W. Metzner, U. Schollwock and K. Schonhammer, 
Phys. Rev. B. {\bf 70}, 075102 (2004).

\bibitem{friedel} J. Friedel, Nuovo Cim. Suppl. {\bf 7}, 287 (1958).

\bibitem{wielen} M. C. M. M. van der Wielen, A. J. A. van Roij and H. van Kempen, Phys. Rev. Lett. {\bf 76}, 1075 (1996).

\bibitem{brazovskii} S. Rouzière, S. Ravy, J.-P. Pouget and S. Brazovski, Phys. Rev. B. {\bf 62}, R16231 (2000).

\bibitem{frid_LL}  R. Egger and H. Grabert, Phys. Rev. Lett. {\bf 75}, 3505 (1995); cond-mat/9604026.

\bibitem{cdw_next} Y. Weiss, M. Goldstein and R. Berkovits, in preparation.

\bibitem{white93} S. R. White, Phys. Rev. B {\bf 48}, 10345 (1993).

\bibitem{g_formula} F. Woynarovich and H. P. Eckle, J. Phys. A {\bf 20}, L97 (1987); 
C. J. Hamer, G. R. W. Quispel, and M. T. Batchelor, ibid. {\bf 20}, 5677 (1987).

\bibitem{shreiber} R. A. Romer and M. Schreiber, Phys. Rev. Lett. {\bf 78}, 515 (1997).  

\bibitem{mahan} G. D. Mahan, {\it Many Particle Physics} (Plenum Press, New York, 1990).



   
\end{thebibliography}
\end{document}